\documentclass[aps,prl,nofootinbib,10pt,tightenlines,notitlepage, twocolumn,floatfix,superscriptaddress,showkeys]{revtex4-2}

\usepackage[utf8]{inputenc}          % file encoding
\usepackage{graphicx}         % Graphics
\usepackage[usenames,dvipsnames]{xcolor}
\usepackage{array,dcolumn,longtable} % Tables
\usepackage{amsmath,amssymb,amsfonts,slashed} % Math symbols
\usepackage{mathtools}          %loads amsmath as well
\usepackage[linktocpage,breaklinks]{hyperref}
\usepackage{mathrsfs}
\usepackage{txfonts}
\usepackage{bm}
\usepackage{stmaryrd}
\usepackage{tensor}
\usepackage[utf8]{inputenc}

\usepackage{epsfig}
\usepackage{epstopdf}

\usepackage{cleveref}

\definecolor{mred}{RGB}{127,0,25}
\definecolor{mdgr}{RGB}{51,51,51}
\definecolor{mag}{RGB}{211, 54, 130}
\definecolor{verm}{RGB}{164, 25, 0}

\hypersetup{colorlinks=true,
            citecolor=NavyBlue,
            linkcolor=NavyBlue,
            urlcolor=NavyBlue}
\usepackage{siunitx}                                  % Parsing numbers and units
\DeclareSIUnit{\fm}{\femto\metre}                     % fm: femtometres

\newcount\hh
\newcount\mm
\mm=\time
\hh=\time
\divide\hh by 60
\divide\mm by 60
\multiply\mm by 60
\mm=-\mm
\advance\mm by \time
\def\hhmm{\number\hh:\ifnum\mm<10{}0\fi\number\mm}

\def\be{\begin{equation}}
\def\ee{\end{equation}}

\def\bfomega{\mbox{\boldmath $\omega$}}
\def\bfOmega{\mbox{\boldmath $\Omega$}}

\def\ve{\varepsilon}
\def\tH{{\tilde H}}
\def\tom{{\tilde \Omega}}

\begin{document}
\title{Particle motion under the conservative piece of the self-force is Hamiltonian}

\author{Francisco M. Blanco}
\affiliation{Department of Physics, Cornell University, Ithaca, NY 14853, USA.}

\author{Éanna É. Flanagan}
\affiliation{Department of Physics, Cornell University, Ithaca, NY 14853, USA.}

%\date{{\color{red}{draft of April 18, 2022; printed \today{} at \hhmm}}}

\begin{abstract}

  We consider the motion of a point particle in a stationary spacetime
  under the influence of a scalar, electromagnetic or gravitational
  self-force. We show that the conservative piece of the first-order
  self-force gives rise to Hamiltonian dynamics, and we derive an
  explicit expression for the Hamiltonian on phase space. Specialized
  to the Kerr spacetime, our result generalizes the Hamiltonian
  function previously obtained by 
  Fujita {\it et.\ al.},
%  , Isoyama, Le Tiec, Nakano, Sago and Tanaka,
  which is valid only
for non-resonant orbits.  We discuss implications for the first law of
binary black hole mechanics.

\end{abstract}

\maketitle
%\iffalse

\vspace{0.1cm}
\noindent \textit{Introduction:} The two body problem in general
relativity has been the focus of intense observational and theoretical
interest in recent years.   On the observational side, LIGO and VIRGO
have detected several dozen coalescences of binary systems containing
black holes and neutron stars \cite{introLIGO1,introLIGO2,introLIGO3}
via the gravitational waves that they emit.
The near future should bring many more 
detections from upgraded instruments, from the next generation ground
based detectors Cosmic Explorer \cite{Evans:2021gyd} and Einstein
Telescope \cite{2010CQGra..27s4002P}, from the space
based detector LISA \cite{Audley:2017drz}, and potentially from pulsar
timing arrays \cite{Antoniadis:2022pcn}.
On the theoretical side a wide variety of approaches valid in
different regimes have been used to understand the dynamics of
black hole binaries with ever greater precision: numerical relativity \cite{Lehner:2014asa}, the 
post-Newtonian approximation \cite{introPN,poissonwill,Levi:2018nxp}, the
post-Minkowskian approximation \cite{Damour:2016gwp} for which amplitude methods from
quantum field theory have been fruitfully brought to bear
\cite{Bern:2021dqo}, the small mass ratio approximation \cite{introEMRI,pound},
and the effective one-body framework which synthesizes information
from the other approaches \cite{Damour:2012mv,Taracchini:2013rva}.

An issue that arises in this field is whether one can define
dissipative and conservative sectors of the dynamics for which the
conservative sector admits a Hamiltonian description.  While this is
not possible in the fully nonlinear, dynamical regime, it has been achieved
in the post-Newtonian and post-Minkowskian approximations to various
orders, and it is a foundational assumption of the effective one body
framework.  Its status within the small mass ratio regime,
however, has been an open question beyond the leading order of geodesic motion.
In that regime the small body is 
treated as a point particle, and the leading order self-force acting
on that body is computed by taking a gradient of a suitably
regularized version of the  body's self field \cite{introEMRI,pound},
computed as a perturbation of the large black hole spacetime.
That force can be split into time-even conservative and time-odd
dissipative pieces.  Hamiltonian descriptions of the conservative motion
have been derived in
special cases (orbits in the Schwarzschild spacetime
\cite{Vines:2015efa} and non-resonant orbits in Kerr \cite{tanaka}).
General orbits in Kerr however have been an open question.

In this Letter we show that the leading order self-forced motion of a
nonspinning body in any stationary spacetime admits a Hamiltonian
description, and derive an explicit expression for the Hamiltonian.
We then discuss a number of applications in the context of black
holes: implications for our  
understanding of the integrability of the motion, a clarification of
the limited domain of validity of the first law of binary black hole
mechanics \cite{LeTiec:2011ab}, and the identification of a new class
of gauge invariant observables that may be useful for comparing
different computational methods.

\bigskip
\noindent{{\em General result in Hamiltonian dynamics:}}  We start by
deriving a general result in the theory of Hamiltonian
systems.  We define a {\it pseudo-Hamiltonian} dynamical system to
consist of a phase space $\Gamma$, a closed,
non-degenerate two form $\Omega_{AB}$ and a smooth pseudo-Hamiltonian function
${\cal H} : \Gamma \times \Gamma \to {\bf R}$, for which the dynamics
are given by integral curves of the vector field
\be
v^A = \Omega^{AB} \frac{\partial}{\partial Q^B} \left. {\cal H}(Q,Q')
\right|_{Q'=Q},
\label{phdef}
\ee
where $\Omega^{AB} \Omega_{BC} = \delta^A_C$ and $Q^A$ are coordinates on $\Gamma$.
%Such systems are Hamiltonian in the special case when
%${\cal H}$ is independent of
%$Q'$.  Another special case is when ${\cal H}$ is symmetric in its
%arguments, so that the system is
%equivalent to a Hamiltonian system with $H(Q) = {\cal H}(Q,Q)/2$.
Pseudo-Hamiltonian systems need not be Hamiltonian, and can be used to
describe dissipation \cite{Galley:2012hx}.

We now specialize to a pseudo-Hamiltonian system which is a
perturbation of a Hamiltonian system, with symplectic form
and pseudo-Hamiltonian
\begin{subequations}
  \label{phpfull}
\begin{eqnarray}
  \Omega_{AB} &=& \Omega_{0\,AB},\\
   {\cal H}(Q,Q') &=& H_0(Q) + \ve {\cal H}_1(Q,Q') + O(\ve^2).
\label{php}
\end{eqnarray}
\end{subequations}
Here $\ve$ is a formal expansion parameter.  We denote by $Q \to \varphi_\tau(Q)$ the
zeroth order Hamiltonian flow, defined by the condition
\begin{equation}
  \left. {d \over d\tau} \right|_{\tau = 0} \varphi^A_\tau(Q)
%  = v_0^A(Q)
= \Omega_0^{AB} \partial_B H_0,
\label{hfz}
\end{equation}
which satisfies the group composition law
\begin{equation}
\varphi_\tau[ \varphi_{\tau'}(Q)] =
\varphi_{\tau+\tau'}(Q).
\label{group}
\end{equation}
The pseudo-Hamiltonian perturbation ${\cal H}_1$ is defined in terms of a function $G :
\Gamma \times \Gamma \to {\bf R}$ via
\be
\label{php1}
{\cal H}_1(Q,Q') = \int_{-\infty}^\infty d\tau' {\tilde
  G}(0,Q,\tau',Q'),
\ee
where we have defined
\begin{equation}
{\tilde G}(\tau, Q, \tau', Q') = G\left[ \varphi_\tau(Q),
  \varphi_{\tau'}(Q') \right].
\label{tildeGdef}
\end{equation}
The function $G$ is assumed to 
satisfy the conditions
\begin{subequations}
  \begin{eqnarray}
\label{condtA}
G(Q,Q') &=& G(Q',Q), \\
\label{condtB}
{\tilde G} (\tau, Q, \tau',Q') &\to& 0 \ \ {\rm
  as} \ \tau\ {\rm or} \ \tau' \to \pm \infty.
  \end{eqnarray}
  \end{subequations}

We now show that with these assumptions, the pseudo-Hamiltonian system
(\ref{phpfull}) is Hamiltonian to linear order in $\ve$.
To do so we need to find a perturbed Hamiltonian $\tH = H_0 + \ve
\tH_1 + O(\ve^2)$, and a perturbed symplectic form $\tom_{AB} =
\Omega_{0\,AB} + \ve \tom_{1\,AB} + O(\ve^2)$, for which the equation of
motion $dQ^A/d\tau = \tom^{AB} \partial_B \tH$ coincides with that
given by Eqs.\ (\ref{phdef}) and (\ref{phpfull}) to $O(\ve)$.  This yields
the requirement
\be
\partial_B \tH_1 - \tom_{1\,BC} \Omega_0^{CD} \partial_D H_0 =
\frac{\partial}{\partial Q^B} \left. {\cal H}_1(Q,Q')
\right|_{Q'=Q}.
\label{condt0}
\ee
We choose the perturbation to the symplectic form to
be\footnote{We originally arrived at this obscure formula by applying
the prescription of Llosa and Vives \cite{llosa1994hamiltonian} for
obtaining Hamiltonians from non-local in time Lagrangians to the
non-local in time action principle for the conservative self-force of
Refs.\ \cite{Galley:2008ih,justin}.}
\be
\tom_{1\,BC} = \left[ {\partial \over \partial Q^B} {\partial \over
    \partial Q^{C'}} \int d\tau \int d \tau'
  \chi(\tau,\tau') {\tilde G}(\tau,Q,\tau',Q')
  \right]_{Q'=Q}
\label{twoform}
\ee
where
\be
\chi(\tau,\tau') = {1 \over 2} {\rm sgn}(\tau) - {1 \over 2}
    {\rm sgn}(\tau').
\label{chidef}
\ee
Because of the antisymmetry property $\chi(\tau',\tau) = -
\chi(\tau,\tau')$ and the symmetry property (\ref{condtA}) of $G$, the
expression (\ref{twoform}) defines a closed two form on phase
space. Using the symplectic form perturbation (\ref{twoform}) and
the pseudo-Hamiltonian perturbation (\ref{php1}) we find that the
requirement (\ref{condt0}) reduces to
\begin{eqnarray}
\partial&&_{B} \tH_1 = \left[{\partial \over \partial Q^B}
\int d\tau' {\tilde G}(0,Q,\tau',Q')
\right]_{Q'=Q}  \nonumber \\
&& +\Omega_0^{CD} \partial_D H_0 \left[ {\partial \over \partial Q^B} {\partial \over
    \partial Q^{C'}} \int d\tau \int d \tau'
  \chi\, {\tilde G}(\tau,Q,\tau',Q')
  \right]_{Q'=Q}.
\label{condt2}
\end{eqnarray}

We now proceed to simplify the second term in Eq.\ (\ref{condt2}), in
several stages.  First, we bring the factor $\Omega_0^{CD} \partial_D H_0$
inside the square brackets and replace it with the tensor
$\Omega_0^{C'D'} \partial_D' H_0$ at $Q'$.  This replacement is valid
because of the subsequent evaluation at $Q'=Q$.  Second, we can
replace the differential operator $\Omega_0^{C'D'} \partial_D' H_0
\partial_{C'}$ using the zeroth order Hamiltonian flow
(\ref{hfz}). The second term becomes
\begin{eqnarray}
  \left\{ \frac{\partial}{\partial Q^B} \right.
    &&
 \left. \frac{d}{d \Delta \tau'} \right|_{\Delta \tau'=0}
 \int d\tau
% \times \nonumber \\ &&\times
\left. \int d \tau'
  \chi\,\tilde G[\tau,Q,\tau',\varphi_{\Delta
    \tau'}(Q')]  \right\}_{Q'=Q}.
\label{second}
\end{eqnarray}
Third, using the definition (\ref{tildeGdef}) of ${\tilde G}$ together with the group property (\ref{group}) of the Hamiltonian flow we have
\be
{\tilde G}[\tau,Q,\tau',\varphi_{\Delta
    \tau'}(Q')]  = 
{\tilde G}(\tau,Q,\tau' + \Delta \tau',Q').
\ee
Hence the term (\ref{second}) can be rewritten as
\be
\left[ {\partial \over \partial Q^B}
  \int d\tau \int d \tau'
  \chi\, {d \over d\tau'}{\tilde G}(\tau,Q,\tau',Q')  \right]_{Q'=Q}.
\label{second1}
\ee
Fourth, we integrate by parts with respect to $\tau'$ and make use of
the condition (\ref{condtB}) to eliminate the boundary terms.  The
derivative of the expression (\ref{chidef}) for the function $\chi$
gives a delta function, $d \chi / d\tau' = - \delta(\tau')$.  The final result is
\be
\left[ {\partial \over \partial Q^B}
  \int d\tau 
  {\tilde G}(\tau,Q,0,Q')  \right]_{Q'=Q}.
\label{second2}
\ee
Using the definition (\ref{tildeGdef}), the symmetry property
(\ref{condtA}) and relabeling $\tau \to \tau'$ this can be written as
\be
\left[ {\partial \over \partial Q^{B'}}
  \int d\tau' 
  {\tilde G}(0,Q,\tau',Q')  \right]_{Q'=Q}.
\label{second3}
\ee
Finally inserting this expression as a replacement for the second term in the condition (\ref{condt2}), we see that the right hand side is now a total derivative, as desired, and the resulting expression for the perturbation to the Hamiltonian is
\be
\label{orig}
\tH_1(Q) = \int d\tau' {\tilde G}(0,Q,\tau',Q).
\ee
This completes the proof that the system (\ref{phpfull}) is Hamiltonian.

We can obtain a more convenient representation of this Hamiltonian
system by making a linearized phase space diffeomorphism parameterized by the
vector field $\ve \xi^A$, under which we have
\begin{subequations}
  \begin{eqnarray}
    \label{newH}
    \tH_1 &\to& H_1 = \tH_1+\mathcal{L}_\xi H_0,\\
    \tom_{1\,AB} &\to& \Omega_{1\,AB} = \tom_{1\,AB} + (\mathcal{L}_\xi \Omega_0)_{AB}.
\end{eqnarray}
\end{subequations}
If we choose $\xi^A = \Omega_0^{AB} \eta_B$ then we find $
  \Omega_{1\,AB} = \tom_{1\,AB} - \partial_A \eta_B + \partial_B
\eta_A$.  We now choose 
\be
\eta_{A} = {1 \over 2} \left[ {\partial \over
    \partial Q^{A'}} \int d\tau \int d \tau'
  \chi\, {\tilde G}(\tau,Q,\tau',Q')
  \right]_{Q'=Q},
\label{etadef}
\ee
which yields from Eq.\ (\ref{twoform}) that $\Omega_{1\,AB}=0$.  Hence the new symplectic form coincides with the unperturbed
symplectic form:
\be
   \Omega_{AB} = \Omega_{0\,AB} + O(\ve^2).
   \label{eq:newcoordsym}
\ee
Similarly by inserting Eq.\ (\ref{etadef}) into Eq.\ (\ref{newH})
%\begin{eqnarray}
%   {\tilde H}_1 &&= H_1 + {1 \over 2}
%\Omega_0^{AB} \partial_A H_0 \times \nonumber \\
%&&\times \Big[{\partial \over
%    \partial \tilde{Q}^{B'}} \int d\lambda \int d \lambda'
%  \chi(\lambda,\lambda') {\tilde G}(\lambda,\tilde{Q},\lambda',\tilde{Q}')
%  \Big]_{\tilde{Q}'=\tilde{Q}}.
%   \end{eqnarray}
%
and simplifying using the same techniques as for
Eq.\ (\ref{condt2}) yields
\be\label{eq:newcoordham}
H_1(Q) = {1 \over 2} \int d\tau' {\tilde G}(0,Q,\tau',Q),
\ee
which differs from the original result (\ref{orig}) by a factor of $2$.

\bigskip
\noindent{{\em Application to motion under the conservative
    self-force:}} We now explain how 
the motion of a
particle under the action of 
its conservative first order gravitational self-force in a stationary spacetime
$(M,g_{ab})$ can be cast as a pseudo-Hamiltonian system of the form
(\ref{phpfull}), by modifying slightly the pseudo-Hamiltonian
construction of Fujita {\it et.\ al.\ }\cite{tanaka}.
This will allow us to apply our Hamiltonian result
(\ref{eq:newcoordsym}) and (\ref{eq:newcoordham}).

For the zeroth order geodesic motion we use phase space coordinates
$(x^\mu, p_\mu)$ with symplectic form $\Omega_0 = d p_\mu \wedge dx^\mu$
and Hamiltonian\footnote{This differs from the Hamiltonian of \cite{tanaka} in that it includes a square root, which is necessary to make $G(Q,Q')$ symmetric in Eq.\ (\ref{Gfn}) below.}
\be
H_0 = - \sqrt{ - g^{\mu\nu}(x) p_\mu p_\nu }.
\label{H0def}
\ee
The time parameter $\tau$ associated with this Hamiltonian is then
proper time normalized with respect to $g_{ab}$, while the conserved
value of $-H_0$ is the mass of the particle. 

For the first order motion, consider a particle at location $x^{\mu'}$
with initial 4-momentum $p_{\mu'}$.  Writing $Q' = (x',p')$, we denote
by\footnote{Our index conventions are unadorned indices for the point $Q =
(x,p)$, primed indices for the point $Q' = (x',p')$, and barred indices for $\varphi_{\tau'}(Q')$.}
$
\varphi_{\tau'}(Q') = [ x^{{\bar \mu}}(\tau'), p_{{\bar \mu}}(\tau')]
$
the geodesic with initial data $Q'$.  From this geodesic we can
compute the Lorenz gauge metric perturbation
$$
h^{\mu\nu}(x;Q') = \frac{1}{\sqrt{ - g^{\mu'\nu'} p_{\mu'} p_{\nu'}}}
\int d\tau'
%\nonumber \\ && \times
G^{\mu\nu\, {\bar \mu}{\bar \nu}}[x, x'(\tau')]
p_{{\bar \mu}}(\tau') p_{{\bar \nu}}(\tau').
$$
Here the symmetric Green's function
$G^{\mu\nu\, {\bar \mu}{\bar \nu}}$ is the average of the retarded and
advanced Green's functions, regularized according to the
Detweiler-Whiting prescription \cite{Detweiler:2002mi,introEMRI}.
The conservative forced motion of the particle is then equivalent at linear order
to geodesic motion in the metric $g_{\mu\nu} + h_{\mu\nu}$, where $Q'$
is held fixed when evaluating the geodesic equation and then evaluated
at $Q'=Q$ \cite{Detweiler:2002mi,pound}.

We can therefore obtain a pseudo-Hamiltonian description of the
dynamics by replacing the metric $g_{\mu\nu}(x)$ in Eq.\ (\ref{H0def})
with $g_{\mu\nu}(x) + h_{\mu\nu}(x,Q')$.  Expanding to linear order in
$h_{\mu\nu}$, comparing with Eqs.\ (\ref{php}), (\ref{php1}) and
(\ref{tildeGdef}), and setting to unity the formal expansion parameter
%\footnote{Whose role will be taken over by the particle mass $m$, see Eq.\ (\ref{hex}) below.}
$\ve$ we can read off the
function $G(Q,Q')$ on phase space to be\footnote{Similar constructions
work for scalar and electromagnetic self-forces.  For a particle
endowed with a scalar charge $q$ and electromagnetic charge $e$ we
replace the initial Hamiltonian expression (\ref{H0def}) with
$-\sqrt{-g^{\mu\nu} (p_\mu - e A_\mu) (p_\nu - e A_\nu)} - q \Phi$.  The expression
(\ref{Gfn}) gets replaced by $-q^2 G_{\rm sc}(x,x')$ in the scalar
case, where $G_{\rm sc}$ is the scalar Green's function, and with
$$
- \frac{ e^2 G^{\mu\mu'}(x,x') p_\mu p_{\mu'}
   }{ \sqrt{-g^{\lambda\sigma} p_\lambda p_\sigma}
  \sqrt{-g^{\lambda'\sigma'} p_{\lambda'} p_{\sigma'}}}
$$
in the electromagnetic case, where $G^{\mu\mu'}$ is the Lorenz gauge
electromagnetic Green's function.}
\be
G(Q,Q') = - \frac{ G^{\mu\nu\,\mu'\nu'}(x,x') p_\mu p_\nu p_{\mu'}
  p_{\nu'} }{2 \sqrt{-g^{\lambda\sigma} p_\lambda p_\sigma}
  \sqrt{-g^{\lambda'\sigma'} p_{\lambda'} p_{\sigma'}}}.
\label{Gfn}
\ee
This function satisfies the symmetry property (\ref{condtA}).
It will also satisfy the decay property (\ref{condtB}) if the
retarded\footnote{The singular Green's function that is subtracted off
in the Detweiler-Whiting regularization prescription does not
contribute here since it vanishes at timelike separations.}
%, and the advanced Green's function will also not contribute.}
Green's function falls off at late times at fixed spatial position.  This is known to be true  
for scalar fields in a class of stationary spacetimes \cite{Hintz:2020roc},
while for black holes it is a lore of the field that perturbations decay at late times as a power law \cite{1998bhp..book.....F}.
This decay was shown for the Weyl scalars in black hole spacetimes by Barack \cite{Barack:1999st},
and it is also generally believed to be true for tensor perturbations, although
it has not yet been established rigorously; see Refs.\ \cite{Dafermos:2017yrz,Dafermos:2021cbw} for
recent developments.
%\footnote{There are homogeneous Lorentz gauge
%solutions in Kerr which grow linearly with time that encode center of
%mass drift \cite{Dolan:2012jg}.  However these should not contribute to the Green's function}.

From this pseudo-Hamiltonian formulation it follows that the motion
under the conservative self-force is
described by the Hamiltonian (\ref{eq:newcoordham}),
in any stationary spacetime for which the retarded Green's function goes to zero
at late times.

\bigskip
\noindent{{\em Specialization to motion near a black
    hole:}} Specialize now to the motion of a particle
orbiting a Kerr black hole.  In this context it is useful to derive an explicit
form for the Hamiltonian in action angle variables.

We use the variables
$(q^\alpha,j_\alpha)=(q^t,q^r,q^\theta,q^\phi,j_t,j_r,j_\theta,j_\phi)$ defined in
Refs.\ \cite{actionangle1,flanagan2}, deformed via
Eq.\ (\ref{etadef}).  In these variables the  
symplectic form is $\Omega = dj_\alpha \wedge dq^\alpha$ and the
full Hamiltonian from Eqs.\ (\ref{eq:newcoordham}) and (\ref{H0def})
is
  \be
  H = H_0(j_\alpha) + H_1(q^\alpha,j_\alpha).
  \label{ham0}
 \ee
The zeroth order geodesic motion is given by $q^\alpha(\tau) = q^\alpha_0 +
\Omega^\alpha_0(j) \tau$, $j_\alpha = $ const, where $\Omega^\alpha_0 = \partial H_0 /
\partial j_\alpha$ are the zeroth order frequencies.

We now fix a value $m$ of the conserved quantity $- H$,
which is the mass of the particle to leading order.
For describing motion on the mass shell $H = -m$
it will be convenient to define rescaled versions of the symplectic
form and Hamiltonian,
\be
      {\hat \Omega}_{AB} = \Omega_{AB}/m, \ \ \ {\hat H} = H /m.
   \label{hatHexpand}
\ee
This rescaling preserves Hamilton's equations.
Using the fact that under the transformation $(x^\mu, p_\mu) \to (x^\mu, s p_\mu)$
with $s > 0$ we have $(q^\alpha, j_\alpha) \to (q^\alpha, s
j_\alpha)$ \cite{flanagan2}, $H_0 \to s H_0$ and $H_1 \to s^2 H_1$
[cf.\ Eq.\ (\ref{Gfn})],
the dynamical system can be written as
\be
   {\hat \Omega} = d J_\alpha \wedge d q^\alpha, \ \ \ 
   {\hat H} = {\hat H}_0(J) + m {\hat H}_1(q,J),
   \label{8d}
   \ee
where $J_\alpha = j_\alpha/m$.
   
Motion on this mass shell can be described in terms a 6 dimensional
Hamiltonian system, which can be derived from the 8 dimensional system
(\ref{8d}) as follows \cite{Arnold}.
Because of the symmetries of the Kerr background the Hamiltonian 
is independent of $q^t$, ${\hat H} = {\hat H}(q^i,J)$
where $q^i = (q^r, q^\theta,q^\phi)$.
Consider paths in the 9-dimensional extended phase space $(q,J,\tau)$ that join an
initial point $(q_1,J_1,\tau_1)$ to a final point
$(q_2,J_2,\tau_2)$. Paths that extremize the line integral of the
Poincar\'e-Cartan one form
%\begin{equation}
$
\int  [J_\alpha dq^\alpha - {\hat H}(q^i,J)d\tau ],
$
 %\label{pc}
%\end{equation}
with $\delta q^\alpha=\delta \tau=0$ at the endpoints, satisfy the
8-dimensional Hamilton equations of motion \cite{Arnold}. We now restrict to paths
lying within the surface ${\hat H}=-1$. Within this surface we can solve
for $J_t=-h(q^i,J_i)$ in terms of the other parameters from the
equation
\be
{\hat H}[q^i, - h(q^i, J_i), J_i ] = -1,
   \label{defh}
\ee
where $J_i = (J_r,J_\theta, J_\phi)$.  The line integral now reduces to
\begin{equation}
    \int \Big[ J_i dq^i - h dq^t\Big] +(\tau_2-\tau_1).
\end{equation}
The second term is a constant and the first term is an extremum under
the variation of paths $[ q^i(q^t),J_i(q^t)]$  that 
connect the two endpoints for which $\delta q^i=0$.
Hence we obtain
%We thus arrive at
%the standard variational principle for
a 6-dimensional Hamiltonian
system with Hamiltonian $h(q^i,J_i)$, time parameter $q^t$ and
symplectic form $d J_i \wedge d q^i$.  
By combining Eqs.\ (\ref{hatHexpand}) and (\ref{defh}) it follows that the
Hamiltonian can be expanded as
\be
h(q^i, J_i) = h_0( J_i) + m h_1(q^i, J_i) + O(m^2).
\label{hex}
\ee
where $h_0$ and $h_1$ are given by ${\hat H}_0(-h_0, J_i) =-1$ and $h_1 = {\hat H}_1(q^i, -h_0, J_i) / \Omega^t_0$.
The zeroth order frequencies are now $\omega^i_0 = \partial h_0 /
\partial J_i = \Omega^i_0/\Omega^t_0$.

The Hamiltonian perturbation $h_1$ is independent of $q^\phi$ due to
the symmetry of the Kerr background, and can be expanded in Fourier
modes\footnote{It is possible to obtain an explicit formula for the
coefficients $h_{1\,{\bf k}}$ starting from a Fourier expansion of the function (\ref{Gfn}) in action angle variables
\begin{eqnarray}
  G(q,J,q',J') &=& \int d\omega \sum_{m, {\bf k}, {\bf k}'} e^{-i \omega( q^t - q^{t'}) -i m (q^\phi - q^{\phi'})}
%  \nonumber \\
%  &&\times 
e^{   i {\bf k} \cdot {\bf q} +i {\bf k}' \cdot {\bf q}' } G_{\omega m {\bf k} {\bf k}'}( J, J'). \nonumber
\end{eqnarray}
Combining this with Eqs.\ (\ref{eq:newcoordham}), (\ref{defh}), (\ref{hex}) and (\ref{hhar}) gives
\be
h_{1\,{\bf k}} = \frac{\pi}{(\Omega^t_0)^2}  \sum_{m, {\bf l}}
G_{\omega m ({\bf k}/2+{\bf l}/2)({\bf k}/2-{\bf
    l}/2)}(J_t,J_i,J_t,J_i),
\nonumber
\ee
where we sum over all pairs of integers ${\bf l} =
(l_r, l_\theta)$ for which $k_r+l_r$ and $k_\theta + l_\theta$ are
even, and we evaluate at $J_t = -h_0(J_i)$ and at $\omega = m
\omega^\phi_0 + ({\bf k} - {\bf l})  \cdot {\bfomega}_0/2$.} on the torus parameterized by ${\bf q} = (q^r, q^\theta)$:
\be
h_1({\bf q}, J_i ) = \sum_{k_r = -\infty}^\infty \sum_{k_\theta = -\infty}^\infty e^{i {\bf k} \cdot {\bf q}} h_{1\, {\bf k}}(J_i).
\label{hhar}
\ee

\bigskip
\noindent{{\em Application: Integrability of dynamics:}} We now turn
to discussing some applications.  Since the motion is Hamiltonian one
can ask whether it is also integrable.  It will be integrable to linear
order if and only if all the resonant mode amplitudes vanish, that is,
\be
h_{1\,{\bf k}}(J_i) =0 \ \ \ {\rm whenever} \ \ \ {\bf k} \cdot
\bfomega_0(J_i) = 0, \ {\bf k} \ne {\bf 0}.
\label{nores}
\ee
This is easy to see, since under a linearized canonical transformation with
generating function $G({\bf q},J_i) = \sum_{\bf k} \exp[i {\bf k} \cdot {\bf q}]
G_{\bf k}(J_i)$ we have
$
h_{1\,{\bf k}} \to h_{1\,{\bf k}} + i ({\bf k}  \cdot \bfomega_0) G_{\bf k}.
$
Thus choosing
$
G_{\bf k}(J_i) = - i h_{1\,{\bf
    k}}(J_i) / {\bf k} \cdot \bfomega_0
$
yields $h_{1\,{\bf k}}=0$ for all nonzero ${\bf k}$ and thus an
integrable system\footnote{The resulting Hamiltonian coincides with
that found by Ref.\ \cite{tanaka}, who excluded resonances.}, and this
choice is possible without divergences only when the condition (\ref{nores}) is satisfied.
Conversely, if the system is integrable there must exist perturbed
versions $J_i + m \delta J_i$ of the action variables which have
vanishing Poisson brackets with the Hamiltonian $h_0 + m h_1$, which
yields at linear order the relation
\be
k_i h_{1\,{\bf k}} = ({\bf k} \cdot \bfomega_0) \delta J_{i\,{\bf k}}
\label{deform}
\ee
between Fourier components, enforcing the condition (\ref{nores}).

An alternative version of the integrability condition
(\ref{nores}) is that the average of
the conservative time derivative of the Carter constant $Q(J_i)$ over any orbit on
any
resonant torus should vanish.  
Computing a time derivative using Eqs.\ (\ref{hex}) and (\ref{hhar}) gives
$dQ/d\tau = \Omega^t_0 (\partial Q / \partial J_i) d J_i / dq^t = - i
\Omega^t_0 (\partial Q / \partial J_i) \sum_{\bf k} k_i h_{1\,{\bf k}}
e^{i {\bf k} \cdot {\bf q}}$.  Now using ${\bf q}(\tau) = {\bf q}_0 +
\bfOmega_0 \tau$, writing the resonant vectors as ${\bf k} = N {\bf
  k}_0 = N(n,-p,0)$ for integers $N$ and taking an orbit average
gives\footnote{We neglect in this calculation the coordinate
transformation (\ref{etadef}), because under $J_i \to
J_i + \delta J_i$ we have ${\dot J}_i \to {\dot J}_i + \omega_0^j
\partial \delta J_i / \partial q^j$ and the resonant Fourier
components of the correction evaluated on a resonant torus vanish.}
\be
\left< {d Q \over d \tau} \right> = - i \Omega^t_0 \left( n {\partial
  Q \over \partial J_r} - p {\partial Q \over \partial J_\theta}
\right)
\sum_{N=-\infty}^\infty N \, h_{1 \, N{\bf k}_0} \, e^{i N q_{\rm res}},
\ee
where $q_{\rm res} = {\bf k}_0 \cdot {\bf q}_0 = n q^r_{0} - p
q^{\theta}_0$ is the resonant combination of the phases.
The left hand side vanishing for all $q_{\rm res}$ is equivalent to
all the resonant amplitudes $h_{1\,N{\bf k}_0}$ vanishing.

One of us conjectured in Ref.\ \cite{flanagan3} that the
linear integrability condition (\ref{nores}) is satisfied in Kerr, based on
the fact that enhanced symmetries present in the
post-Newtonian limit enforce this condition.  However, this was a weak
argument, since it is possible for symmetries to be present only near
the boundary of phase space that corresponds to the post-Newtonian
limit, and not in the interior (just as for asymptotic spacetime symmetries).
Indeed, recently Nasipak and Evans have shown numerically that $\left<
dQ/d\tau \right>=0$
fails for conservative scalar
self-forces in Kerr on resonances
\cite{integrability1,integrability2}.  The gravitational self-force
case is presumably similar, although this will need to be confirmed
numerically (see Ref.\ \cite{vandeMeent:2017bcc}).

%Assuming that the integrability condition (\ref{}) fails, 
If the gravitational case is indeed non-integrable, 
the qualitative consequences for the conservative dynamics are well
understood in general contexts from the theory of weakly 
perturbed Hamiltonian systems \cite{Arnold,RevModPhys.56.737}.  They
have been explored in the contexts 
of tidal and other perturbations to extreme mass ratio inspirals in
Refs.\ \cite{Apostolatos:2009vu,Bonga:2019ycj,Lukes-Gerakopoulos:2021ybx,Bronicki:2022eqa}.
Suppose we focus attention on one resonant torus  $J_i = J^*_{i}$ and neglect the effect of other resonances.
First, away from this torus the invariant tori $J_i = $ constant
are deformed [cf.\ Eq.\ (\ref{deform})] but preserved (as
predicted by the KAM theorem \cite{Arnold}).  Second, within a shell of width $J_i -
J^*_i \sim \sqrt{m}$ the dynamics is altered: In the $m \to 0$ limit
the resonant torus is destroyed and replaced by a number of islands of
size $\sim \sqrt{m}$ in
phase space within which the  motion is integrable\footnote{This can
be seen explicitly in the description of the near-resonance dynamics
derived by  van de Meent, Eq.\ (18) of Ref.\ \cite{vandeMeent:2013sza}, dropping
the  dissipative terms (the first term on the right hand side and half
of the oscillatory terms); the solutions consist of rotational or
librational (islands) motions, depending on the energy.} \cite{Bronicki:2022eqa}.  One  can define 
action angle variables within each island, but they do not join
continuously onto the global action angle variables.
%away from the
%resonance.
At finite $m$ chaotic regions develop within the shell.
Third, motion that starts within the shell is confined to remain within it by the
surrounding surviving invariant tori, since the system is effectively
two dimensional ($J_\phi$ is conserved)
\cite{RevModPhys.56.737}. There are no large excursions to $J_i - J^*_i
\sim O(1)$,  % (Arnold diffusion)
%that are generic in higher dimensions.
unlike in higher dimensions.

When one considers the full $O(m)$ dynamics with the dissipative component of
the self force included, the non-integrable mode coefficients $h_{1 \,
  {\bf k}}$ can drive transient resonances which give $O(\sqrt{m})$
kicks to the action variables $J_i$ \cite{flanagan3}, and also
sustained resonances in which the orbit evolves along
a non-adiabatic path in the space of parameters $J_i$ maintaining the condition
${\bf k} \cdot \bfomega_0(J_i)=0$ \cite{vandeMeent:2013sza}.  However
neither of these are smoking gun  signatures of the breakdown of
integrability, since both can be produced when $h_{1\,{\bf k}}=0$ by
the oscillatory dissipative  components of the  self force
\cite{vandeMeent:2013sza}.

Non-integrability would also complicate the dynamics away from the
resonant islands in phase space. 
If one computes the dynamics using the linear prescription described
after Eq.\ (\ref{nores})
for eliminating the oscillatory terms in the Hamiltonian (\ref{hex}), ignoring the
divergences, the
resulting fractional errors caused by the nearest
strong resonance scale as $\sim m^2 | h_{1{\bf
    k}}|^2  (J-J^*)^{-4}$.
It is possible to achieve smaller errors
$\sim m^3 | h_{1{\bf k}}|^3  (J-J^*)^{-6}$ by
using a second order canonical transformation to eliminate the
oscillatory terms in (\ref{hex}) from $h_1$ through $O(m^2)$,
%(like the transformations used in Ref.\ \cite{VanDeMeent:2018cgn}),
at the 
price of a more complicated description of the dynamics.
In either case the errors become of order unity in the vicinity of the
resonant islands.

\bigskip
\noindent{{\em Application: First law of binary
    black hole mechanics:}} In the absence of resonances, our
Hamiltonian (\ref{ham0}) directly 
yields a version of the first law, as in Ref.\ \cite{tanaka}.
%whose derivation is similar
%to that of Ref.\ \cite{tanaka} but simpler due to the modified initial
%Hamiltonian (\ref{H0def}).
We eliminate all $q$ dependent terms in
(\ref{ham0}) using a canonical transformation as described after
Eq.\ (\ref{nores}). We regard $H$ as a function $H = H(j_\alpha,
M_{\rm irr}, S_{\rm bh})$ of the action variables $j_\alpha$ and of the
irreducible mass $M_{\rm irr}$ and spin $S_{\rm bh}$ of the large black hole.  Taking a
variation and using $H = - m$ gives
\be
- \delta m = \Omega^t \delta j_t + \Omega^i \delta j_i +
\frac{\partial H}{\partial M_{\rm irr}} \delta M_{\rm irr}
+ \frac{\partial H}{\partial S_{\rm bh}} \delta S_{\rm bh},
\ee
where $\Omega^\alpha = \partial H / \partial j_\alpha$ are the
frequencies accurate to subleading order in $m$.
Identifying $-j_t$ as the orbital energy $E$, dividing by $\Omega^t$,
and adding the variation of the background black hole mass $M_{\rm bh}(M_{\rm irr}, S_{\rm bh})$
gives
\be
\delta(M_{\rm bh}+ E) = z \delta m + \omega^i \delta j_i + z_{\rm bh}
\delta M_{\rm irr} + \Omega_{\rm bh} \delta S_{\rm bh}.
\label{FLBM}
\ee
Here $z  = 1 / \Omega^t$ is the redshift invariant, $\omega^i =
\Omega^i / \Omega^t$, $z_{\rm bh} = \partial M_{\rm bh} / \partial M_{\rm irr} + z \partial H / \partial M_{\rm irr}$ and $\Omega_{\rm bh} = \partial M_{\rm bh} / \partial S_{\rm bh} + z \partial H / \partial S_{\rm bh}$.
Equation (\ref{FLBM}) yields a form of the
first law for binaries\footnote{Equation (\ref{FLBM}) is not quite the
conventional form of the first law beyond the leading order in $m$.  The conventional form would require the quantity
$M_{\rm bh} + E$ on the left hand side to coincide with the Bondi mass to $O(m^2)$,
whereas it is known to coincide only to $O(m)$ \cite{Pound:2019lzj}.
Additionally we have identified the on-shell value of the Hamiltonian
with minus the particle mass $m$, and this
relation can have a correction to subleading order in $m$.
Nevertheless, our form is sufficient to illustrate the difficulties
caused by non-integrability, which are generic for all forms of
the first law.} to subleading order in $m$.

This derivation of the first law required the integrability assumption (\ref{nores}).
We now explain how the first law would break down if that assumption
is violated as discussed above.
%Our Hamiltonian (\ref{ham0}) now allows us to identify how the first law breaks
%down because of the presence of resonances and the nonintegrability
%discovered by Refs.\ \cite{integrability1,integrability2}.
The first law requires a
labeling of time-averaged orbits by some smooth set of parameters.
%The choice of action variables $j_i$ yields parameters that diverge at resonances, as discussed
%around Eq.\ (\ref{nores}) above.  While this difficulty might be
%remedied by choosing a different set of parameters, a more serious
%problem is the fact that the
However, when (\ref{nores}) is violated integrable motions near a resonance fall
into different types that are disconnected from one another.  Within
an island one can define new action angle variables ${\tilde q}^i,
{\tilde j}_i$, but these cannot join smoothly onto the deformed
action-angle variables outside
the islands.  Thus, the best one can hope for is set of distinct first laws,
one for each disconnected component of integrable motion.
Also the number of such components is formally infinite, since
the resonances are dense in phase space.  In practice
only the few 
resonances for which the order $|k_r| + |k_\theta|$ is not large will
be significant: the width of an island scales as $\sim \sqrt{m} | h_{1 {\bf
    k}}|$ \cite{vandeMeent:2013sza} which will go 
exponentially to zero as the order
increases, assuming the Hamiltonian is a smooth
function on the torus.

\bigskip
\noindent{{\em Application: Gauge invariant observables:}} Gauge
invariant observables such as invariant redshifts, frequencies 
of innermost stable circular orbits, etc. have proven enormously
useful for cross checks between different computational methods
\cite{pound}. The simple form (\ref{eq:newcoordham}) of our Hamiltonian may
be helpful for computing such observables, since one expects the
complicated phase space coordinate transformation (\ref{etadef}) 
not to be relevant for gauge invariant observables.

For generic orbits, non-integrability of the dynamics would
impede the definition of such observables.  For example one can no
longer label orbits by their three fundamental frequencies of motion.
However new gauge invariant observables do arise in this context,
the resonant amplitudes $h_{1 {\bf k}}$ themselves,
for which the action-angle variables are defined geometrically
at zeroth order \cite{flanagan2}
and which at first order are invariant under linearized phase space coordinate transformations.
These observables are not accessible from within post-Newtonian or
post-Minkowski theory, but could be useful for comparisons between self-force
theory and numerical relativity.

\bigskip
\noindent{{\em Conclusions:}} We have shown that the conservative
dynamics of two body problem in general relativity in the small mass
ratio limit is Hamiltonian to the first subleading order, when the
small body is
nonspinning.
%In this limit Nasipak and Evans have shown that the dynamics is
%non-integrable \cite{integrability1,integrability2}.
It would be
interesting to extend this result to include the spin of
the small body, and to second-order conservative self-forces.

%%%%%%%%%%%%%%%%%%%%%%%%%%%%%%%%%%%%%%
%.  Acknowledgments
%%%%%%%%%%%%%%%%%%%%%%%%%%%%%%%%%%%%%%
\vspace{0.1cm}
\noindent \textit{Acknowledgments:} We thank Adam Pound, Leo Stein,
Justin Vines and Neils Warburton for helpful discussions. This
research was supported in part by NSF grant PHY-2110463.

\color{black}

%%%%%%%%%%%%%%%%%%%%%%%%%%%%%%%%%%%%%%
%.  Bibliography
%%%%%%%%%%%%%%%%%%%%%%%%%%%%%%%%%%%%%%
%\bibliographystyle{}
\bibliography{Ref}

\end{document}